\documentclass[journal,twoside,web]{ieeecolor}
\usepackage{tmi}
\usepackage{cite}
\usepackage{textcomp}

\usepackage{amsmath,amsfonts,amssymb}
\usepackage{algorithm,algpseudocode}
\usepackage[nodisplayskipstretch]{setspace}

\usepackage{graphicx}
\usepackage{multirow}
\usepackage{colortbl}
\usepackage{makecell}
\usepackage{amsmath}
\usepackage{url} 

\makeatletter
\let\NAT@parse\undefined
\makeatother

\usepackage{graphicx}
\usepackage{multirow}
\usepackage{xcolor,colortbl}
\usepackage{caption}
\usepackage{subcaption}
\definecolor{tmi_blue}{cmyk}{100,0.37,0.0,0.15}
\usepackage{hyperref}
\hypersetup{
    colorlinks=true,
    linkcolor=tmi_blue,
    citecolor=tmi_blue,
    urlcolor=tmi_blue
}

\def\BibTeX{{\rm B\kern-.05em{\sc i\kern-.025em b}\kern-.08em
    T\kern-.1667em\lower.7ex\hbox{E}\kern-.125emX}}
\begin{document}
\title{XReal: Realistic Anatomy and Pathology-Aware X-ray Generation via Controllable Diffusion Model
}

%
\author{Anees Ur Rehman Hashmi, Ibrahim Almakky, Mohammad Areeb Qazi, Santosh Sanjeev, Vijay Ram Papineni, Jagalpathy Jagdish, Mohammad Yaqub,
\thanks{
(Corresponding author: Ibrahim Almakky.)}%
\thanks{Anees Ur Rehman Hashmi, Ibrahim Almakky, Mohammad Areeb Qazi, Santosh Sanjeev, and Mohammad Yaqub are with Mohamed bin Zayed University of Artificial Intelligence, Abu Dhabi, UAE (email: firstname.lastname@mbzuai.ac.ae).}
\thanks{Vijay Ram Papineni and Jagalpathy Jagdish are with Sheikh Shakhbout Medical City, Abu Dhabi, UAE.}
}

\maketitle

\begin{abstract}

Large-scale generative models have demonstrated impressive capabilities in producing visually compelling images, with increasing applications in medical imaging. However, they continue to grapple with hallucination challenges and the generation of anatomically inaccurate outputs. These limitations are mainly due to the reliance on textual inputs and lack of spatial control over the generated images, hindering the potential usefulness of such models in real-life settings. In this work, we present XReal, a novel controllable diffusion model for generating realistic chest X-ray images through precise anatomy and pathology location control. Our lightweight method comprises an Anatomy Controller and a Pathology Controller to introduce spatial control over anatomy and pathology in a pre-trained Text-to-Image Diffusion Model, respectively, without fine-tuning the model. XReal outperforms state-of-the-art X-ray diffusion models in quantitative metrics and radiologists' ratings, showing significant gains in anatomy and pathology realism. Our model holds promise for advancing generative models in medical imaging, offering greater precision and adaptability while inviting further exploration in this evolving field. The code and pre-trained model weights are publicly available at \href{https://github.com/BioMedIA-MBZUAI/XReal}{https://github.com/BioMedIA-MBZUAI/XReal}.

\end{abstract}

\begin{IEEEkeywords}
Diffusion Model, \and Clinical Realism, \and Image Generation, \and X-ray
\end{IEEEkeywords}

\begin{figure*}[t]
    \centering
    \includegraphics[width=\textwidth]{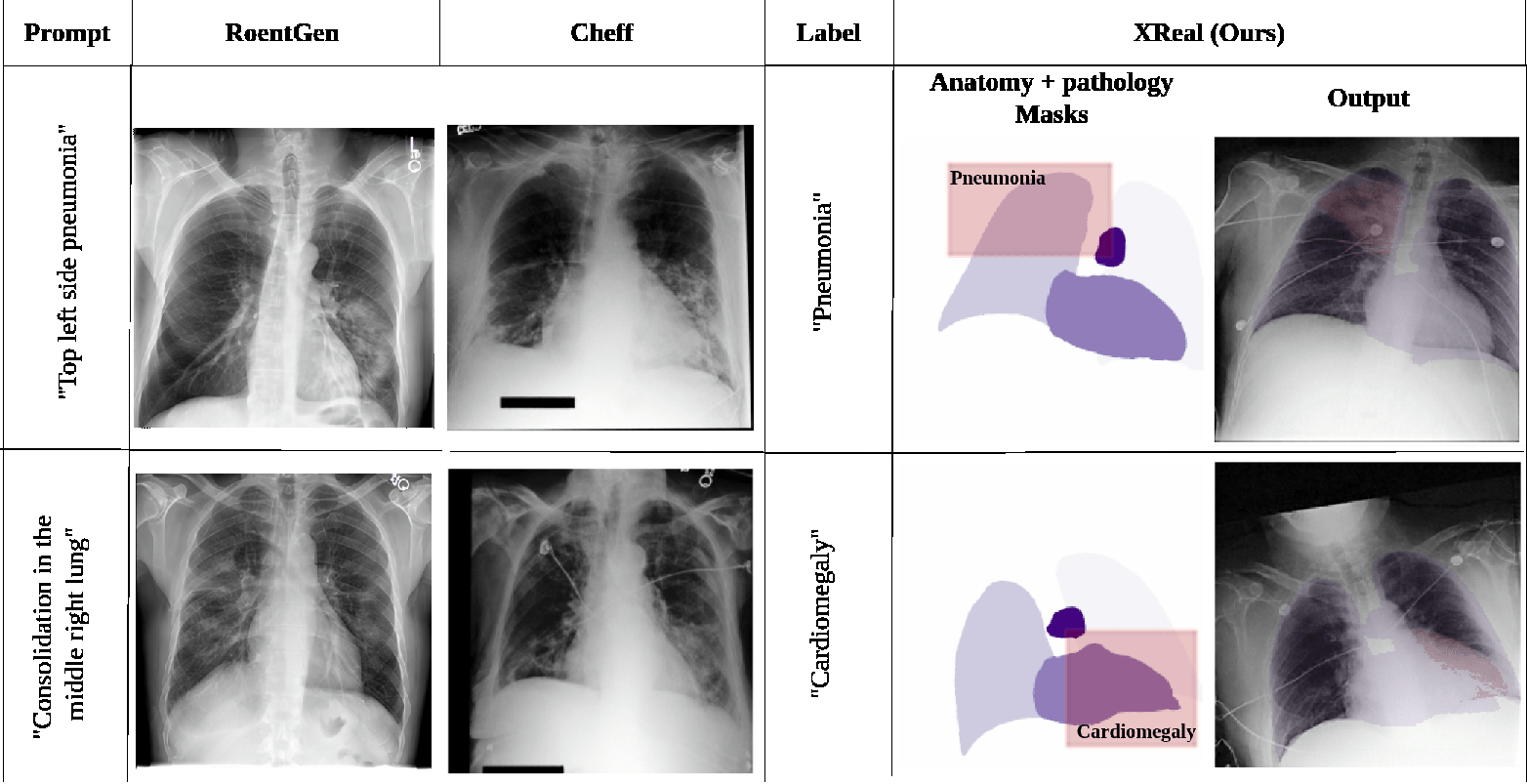}
    \caption{X-ray generation using different diffusion models. As text-to-image models, RoentGen \cite{chambon2022roentgen} and Cheff \cite{weber2023cascaded} struggle to follow the pathology location information specified in the prompts and do not offer any anatomy control. Our proposed XReal model provides precise control over both anatomical and pathology manifestations through the use of input segmentation masks, significantly enhancing the clinical realism of generated X-ray images.}
    \label{fig:result}
\end{figure*}

\section{Introduction} \label{sec:intro}

Deep generative models have shown remarkable success in many applications, including healthcare, with the ability to generate high-quality text and images with intricate details \cite{ramesh2022hierarchical,brown2020language,rombach2022high}. However, despite significant advancements in image quality, these models frequently struggle with hallucinations, leading to the generation of images containing illogical and unrealistic content \cite{rawte2023survey}. One primary factor contributing to this challenge is their reliance solely on textual input for conditioning, which often falls short of providing complete guidance for logical and realistic image generation \cite{avrahami2023spatext,casanova2023controllable}.

Text-to-image generative models, including Variational AutoEncoder (VAEs) \cite{kingma2013auto}, Generative Adversarial Networks (GANs) \cite{goodfellow2014generative}, and more recently, diffusion models \cite{dhariwal2021diffusion} have shown promising generative capabilities for high-quality image synthesis in the medical domain  \cite{gu2023biomedjourney,chambon2022roentgen,weber2023cascaded}. However, relying solely on free-form text to generate images \cite{chambon2022roentgen,weber2023cascaded} limits the control over critical spatial information in medical images, especially affecting anatomical structures and pathology manifestations. Fig.\ref{fig:result} depicts this issue in the text-to-image models that struggle to follow the spatial information provided in the text prompt. The absence of spatial control in these models makes it almost impossible to control the fine details of the organs and diseases in the generated images. Furthermore, it is also very important to control the relative location of the disease manifestation and the organs because many diseases are plausible only when manifested in a specific location relative to the organs of interest in the body (e.g., cardiomegaly and heart). This concern is particularly amplified in chest X-ray images where a particular disease can manifest in many regions simultaneously (e.g., bilateral pneumonia), and minor alterations to its manifestation in the generated image can significantly impact the disease identification and overall image interpretation. Hence, the absence of spatial control in generative models affects the clinical realism of the generated data and limits their practical applications in the medical domain (e.g., for radiologist training).

To address this, we introduce XReal, a diffusion model capable of generating high-quality, clinically realistic X-ray images with control over anatomy and pathology and pathology manifestation. Through spatial control, our lightweight model generates X-ray images, enhancing the usefulness of the generated data for downstream medical applications. To this effect, the main contributions of this work are as follows:

\begin{itemize}
    \item We introduce XReal, a novel pathology and anatomy-aware controllable diffusion model for realistic X-ray image generation. XReal can generate high-quality and clinically realistic X-ray images with precise control over the organs’ location, size, shape, and pathology manifestation.

    \item We conduct extensive experiments, comparing XReal with existing image generation models and demonstrate state-of-the-art performance using a combination of quantitative metrics and expert radiologists' evaluation. 
\end{itemize}

\section{Related Work}

\subsection{X-Ray Generation}
 
Several works have been proposed for synthetic CXR generation. GANs have been a widely used type of generative model as they offer high fidelity and fast sampling. However, GAN training is highly unstable due to its adversarial design and often faces problems like mode collapse, resulting in a lack of image diversity. Previously, \cite{segal2021evaluating} used the progressive-growing GAN (PGAN \cite{karras2017progressive}) for class-guided X-ray synthesis. \cite{ng2023generative} used the Deep Convolutional GAN (DCGAN \cite{radford2015unsupervised}) and the Wasserstein GAN with Gradient Penalty (WGAN-GP \cite{gulrajani2017improved}) to augment data for X-ray classification. On the other hand, \cite{ciano2021multi} generated X-rays conditioned on organ segmentation masks using a multi-stage GAN. \cite{yang2020xraygan} proposed the XRayGAN framework to generate multi-view X-ray images using clinical reports.

More recently, diffusion models have been introduced for X-ray synthesis due to their ability to produce higher quality and more diverse images \cite{dhariwal2021diffusion}. \cite{chambon2022adapting} provides one of the first works on adapting a pre-trained stable diffusion model \cite{rombach2022high} for medical report-to-X-ray synthesis. Their work shows the effect of using out-of-domain pre-trained VAE and text-encoder and textual inversion to learn new medical concepts in few-shot learning. Similarly, \cite{chambon2022roentgen} investigated the impact of different strategies to adopt the stable diffusion \cite{rombach2022high} architecture for X-ray generation. Their study showed that fine-tuning both the U-Net and CLIP (Contrastive Language-Image Pre-Training \cite{radford2021learning}) text encoder in stable diffusion yields the highest image fidelity and conceptual correctness. \cite{weber2023cascaded} trained a cascaded diffusion model for the report to X-ray generation task. Their model incorporates two stages: one for text-to-image generation and another to enhance the resolution of the initial image to high resolution. This two-stage approach enables high-resolution image generation with reduced computational requirements by keeping the text-to-image model lightweight and using the second stage to upscale the initial output. Another study \cite{packhauser2023generation} used the Latent Diffusion Model (LDM) to generate class-conditional X-ray images and employed a privacy-enhancing sampling strategy to ensure the non-transference of biometric information during the image generation process. Although textual conditioning or class labels-based X-ray generation remains an active research area, the utilization of spatial information, particularly concerning anatomy and X-ray pathologies, remains largely unexplored.

\subsection{Spatial Control in Diffusion Models}

Prior research on guiding diffusion models with spatial input has predominantly focused on natural image generation, with little focus on medical images. Within this domain, three main strategies have emerged for incorporating spatial control into diffusion models.

\begin{figure*}[t]
    \centering
    \includegraphics[width=\textwidth]{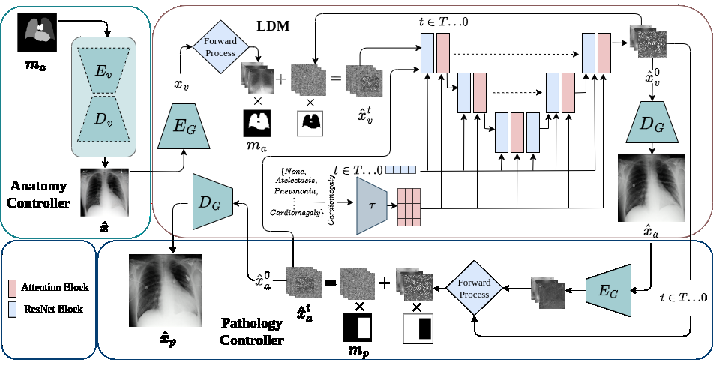}
    \caption{XReal has three components: 1) Anatomy Controller, 2) Latent Diffusion Model, and 3) Pathology Controller. It uses a two-stage process to generate the final image $\hat{x}_p$. The Anatomy Controller guides the LDM to generate image $\hat{x}_a$ based on the anatomy mask $m_a$ without using any textual input (text = ``'' or None). The Pathology Controller infuses the pathology $p$ (text = $p$) into $\hat{x}_a$ at $m_p$ to obtain the final image $\hat{x}_p$.}
    \label{fig:method_overview}
\end{figure*}

The first approach involves training a diffusion model tailored to a specific task. This necessitates access to a substantial paired mask, image-to-image dataset. For instance, \cite{liu2023textdiff} leveraged text and mask modules to achieve image super-resolution using a diffusion model. Another study \cite{meng2022sdedit} utilized a partially noisy input image to condition the diffusion model. Additionally, \cite{xie2023smartbrush} trained a diffusion model to inpaint objects using both text and shape guidance. In the medical imaging domain, \cite{van2023echocardiography} generated the video echo-cardiographs through semantic map guidance of the diffusion model. These semantic maps are added directly to the decoder of the 3D UNet diffusion model to add spatial conditions. Another study in the medical domain \cite{dorjsembe2023conditional} employed 3D segmentation masks to generate MRI volumes using a diffusion model. One significant limitation of this approach is the requirement for a substantial amount of paired data, which is very scarce in the medical domain. A major drawback of these methods for diffusion models is their inflexibility, which results from task-specific training. Once adapted for a single task, such methods require the entire diffusion model to be retrained for each new application or dataset.

The second approach focuses on manipulating the cross-attention mechanism in pre-trained diffusion models \cite{wu2023harnessing,couairon2023zero}. For instance, \cite{park2022shape} used shape masks to decouple irrelevant attention in text-to-image diffusion. Similarly, \cite{xie2023boxdiff} employed bounding boxes to constrain cross-attention within the stable diffusion model. While lightweight, attention-based methods are highly sensitive to textual input and rely heavily on text-image interactions, making precise spatial control challenging. Additionally, these methods cause a drop in image quality by introducing artifacts or unintended distortions in the generated images. The influence of attention mechanisms on the diffusion models can be intricate, and improper manipulation can lead to undesirable artifacts in the final output.

The third approach is based on hyper-networks, which are smaller networks used to guide the output of a larger model. In generative models, hyper-networks guide the internal image representation in larger models for specific image manipulations while keeping their original image generation capabilities intact. This enables the adaption of large models for specific purposes and introduces specific conditioning without re-training the large models. \cite{zhang2023adding} introduces ControlNet, which uses the pre-trained UNet encoder of the diffusion model as a guiding network to steer the frozen diffusion model. This enables spatial control over the generated images without retraining the diffusion model. However, previous research \cite{ratzlaff2019hypergan, kumar2023coronetgan} in this domain has primarily focused on introducing control within the context of natural image generation, with limited attention to medical image generation. Furthermore, there has been no prior attempt to add spatial control over the generation of X-ray images and their associated lesions. 
\section{Method} 
\label{sec:metehod}

We propose XReal to generate an image $\hat{x}_p$ given an anatomy mask $m_{a}$, a pathology mask $m_{p}$, and the pathology label $p \in P$, where $P = \{p_1, \dots p_n\}$ is the set of $n$ possible pathologies. The generated image $\hat{x}_p$ should follow the anatomical structure specified in $m_{a}$ while manifesting the pathology $p$ in the specified location within $m_p$. As depicted in Fig. \ref{fig:method_overview}, XReal consists of an Anatomy Controller component followed by a Latent Diffusion Model (LDM) and a Pathology Controller. In the following subsections, we describe how these components work together to achieve the final generative outcomes.

\subsection{Anatomy Controller}

To control the anatomical structure of the generated image, $\hat{x}_p$, we developed an Anatomy Controller consisting of a VAE comprising of an encoder $E_v$ and a decoder $D_v$. The Anatomy Controller is trained to take a segmentation mask of anatomical structures $x_a$ as input and generate an X-ray image $\hat{x}$. Therefore, the Anatomy Controller VAE is trained to generate $\hat{x}$ as follows: $\hat{x} = D_v(E_v(m_a))$, where $\hat{x} \approx x$ and $(x, m_a)$ are an X-ray image and its corresponding input anatomy mask, respectively. The $\hat{x}$ looks similar to an X-ray image with the overall anatomical structure as provided in $m_a$ but does not have any fine-grained X-ray image details and has low image quality. This X-ray image, $\hat{x}$, generated through the Anatomy Controller, is used to infuse spatial information into the pre-trained diffusion model in the subsequent steps. This is possible due to the property of the VAE's latent space, which preserves the structural information of the input. 

\subsection{Latent Diffusion Model}
Diffusion models \cite{ho2020denoising} are probabilistic generative models that generate an image through iterative denoising of noisy inputs. The training process of diffusion models involves the addition of Gaussian noise to a clean image over a series of $T$ timesteps. Following this, the model learns to denoise the noisy image in the backward diffusion process, gradually removing the noise and recovering the original image. While diffusion models can generate high-quality and diverse images, the backward process requires iteration over a large number of timesteps ($T$), making them computationally expensive. Alleviating this computational cost, we adopt the Latent Diffusion Model (LDM) \cite{rombach2022high}, where the diffusion process is applied in a latent space. LDM comprises of a pre-trained VAE \cite{kingma2013auto} consisting of an encoder $E_G$ and decoder $D_G$ and a text-to-image diffusion model in its latent space. 

In this work, we employ a VAE trained for image-to-image reconstruction tasks for our LDM. In such a manner, the VAE encoder $E_G$ encodes the output of the Anatomy Controller as follows: $\hat{x}_v = E_G(\hat{x})$.

\begin{figure}[t]
    \centering
    \includegraphics[width=0.5\textwidth]{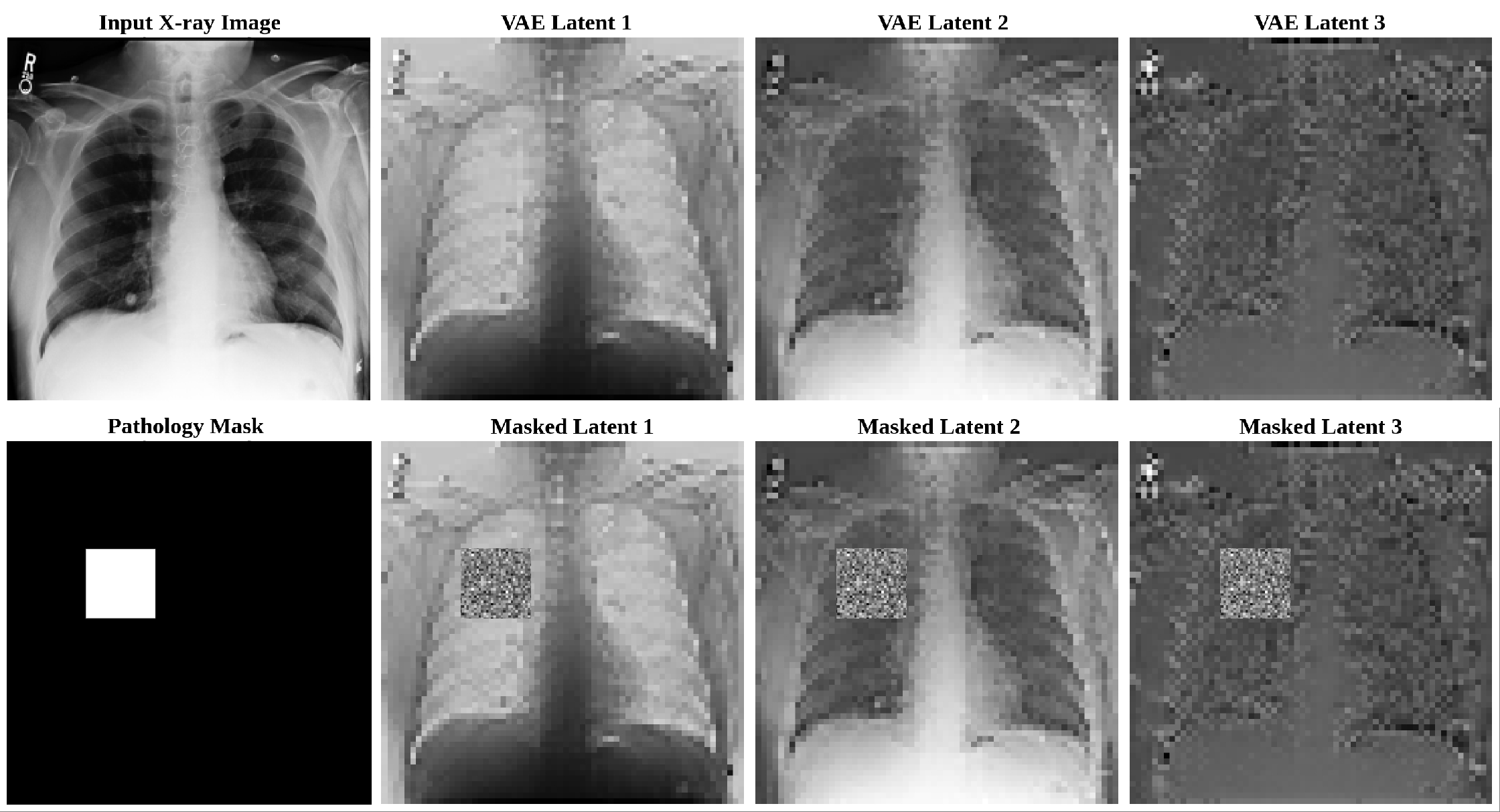}
    \caption{The top row shows the Latent space of VAE in LDM. The VAE encoder, $E_G$, preserves the anatomy of the input X-ray image in the latent space, which can be manipulated to provide spatial control. The bottom row has a sample pathology mask $m_p$ for pneumonia. The pathology controller combines this $m_p$ with latents of the X-ray image to add a specific pathology $p$. }
    \label{fig:vae_latent_space}
\end{figure}

After this, $\hat{x}_v$ and $m_a$ are infused in order to introduce anatomical guidance to the latent image representation of $E_G$. The latent $\hat{x}_v$ still maintains the structural features of the input X-ray image as shown in Fig. \ref{fig:vae_latent_space}. We make use of this spatial information to guide the diffusion model by adding Gaussian noise ($x_T$) to $\hat{x}_v$ to get $\hat{x}^t_v$ through the forward diffusion process as described in \cite{ho2020denoising}. This noisy latent representation obtained by the forward diffusion process is then combined with $\hat{x}^{t-1}_v$ and $m_a$ using the following equation. 

\begin{equation} \label{eq:anat_mask}    
    \hat{x}^t_v = \hat{x}^t_v \times m_a + (1-m_a) \times \hat{x}^{t-1}_v
\end{equation}

where $\hat{x}^{t-1}_v$ is the output of the diffusion model from the previous timestep and is initialized as sampled Gaussian noise ($\epsilon$) when $t = T$

The final noisy latent representation, $\hat{x}^t_v$, is passed to the diffusion model $G$ to generate $\hat{x}^0_v$. Equation \ref{eq:anat_mask} allows LDM's UNet, $G$, to utilize the anatomical information present in the noisy $\hat{x}^0_v$ while generating the peripheries (clavicle, humerus, head, etc.) using the random Gaussian noise $\epsilon$. Our empirical analyses show that the initial backward steps in the denoising process determine the overall structure of the generated image, while later steps enhance the structure. Motivated by this, we apply anatomical guidance for initial $s$ out of $T$ backward diffusion steps. Finally, after $T$ iterations, the LDM's VAE decoder $D_G$ decodes $\hat{x}^0_v$ to $\hat{x}_a$ image containing the desired anatomy.

\begin{algorithm}[t]
\caption{XReal inference process with anatomy and pathology control}
\label{alg:main_algo}
\begin{algorithmic}
\Require $\hat{x}_v \in  \mathbb{R}^{3\times64\times64}$ \Comment{Latent of $\hat{x}$}
\Require $m \in  \mathbb{R}^{3\times64\times64}$ \Comment{Anatomy or Pathology Mask}
\Require $s \in \mathbb{R}$ \Comment{Number of steps to mask for}
\Require $p$ \Comment{Pathology label}

\State $\hat{x}^{t-1}_v \sim \mathcal{N}(0, I)$

    \For {t = T,...,0}
    
        \setstretch{1.1}

        \If{$t\geq (T-s)$}
        
            $\epsilon \sim \mathcal{N}(0, I)$ 
            
            $x^t_v = \sqrt{\bar{\alpha_t}\hat{x}_v} + \sqrt{1- \bar{\alpha_t}}\epsilon $\Comment{Forward process}
            
        
            $\hat{x}^{t}_v = m \cdot x^t_v + (1-m) \cdot \hat{x}^{t-1}_v$     
        
        \EndIf
        
        $z \sim \mathcal{N}(0, I)$ if $t > 1$ else $z = 0$

        $\hat{x}^{t-1}_v = \frac{1}{\sqrt{\alpha_t}} (\hat{x}^{t}_v - \frac{1-\alpha_t}{\sqrt{1-\bar{\alpha_t}}} \epsilon_\theta(\hat{x}^{t}_v, t, p) ) + \sigma_t z$
        
        \setstretch{1}
        
    \EndFor  \\
    
\Return $x_0$

\end{algorithmic}
\end{algorithm}

\subsection{Pathology Controller} 

Given an image $\hat{x}_a$ generated by Anatomy Controller, the input pathology mask $m_p$ and the pathology $p$, our Pathology Controller generates the image $\hat{x}_p$ containing $p$ at $m_p$ while preserving the anatomy $m_a$ in $\hat{x}_a$. We use the inpainting capabilities of the text-to-xray diffusion model $G$ and fill the $m_p$ region in $\hat{x}_a$ with pathology $p$. In a similar way to our Anatomy Controller, we first encode $\hat{x}_a$ to the latent space using $E_G$ of LDM. Followed by the addition of random Gaussian noise ($\epsilon$) to the input image $\hat{x}_a$ using the forward diffusion process that yields $\hat{x}^t_a$. Using pathology mask $m_p$, we then combine $\hat{x}^{t}_a$ and the output of the diffusion model from the previous timestep $\hat{x}^{t-1}_a$ (initially set to Gaussian noise $\epsilon$) such that $\hat{x}^{t-1}_a$ is overlayed at the location where we want to put the pathology $p$. The rest of the X-ray image remains similar to  $\hat{x}_a$, preserving the anatomy as shown in Fig. \ref{fig:vae_latent_space} (bottom). The Markovian chain process of the diffusion model makes the input image $\hat{x}_a^t$ as a prior for $\hat{x}_a^{t-1}$. Thereby generating the pathology smoothly by using the existing information in the input and avoiding unrealistic artifacts. 
We combine the output from the diffusion model $\hat{x}^t_p$ with the noisy version of the $\hat{x}_a$ as follows.

\begin{equation} \label{eq:path_mask}    
    \hat{x}^t_a = \hat{x}^{t-1}_a \times m_p + (1-m_p) \times \hat{x}^t_a
\end{equation}

We repeat this iterative masking and denoising process for $T$ timesteps to obtain $\hat{x}^0_a$, decoded by $D_G$ to the final X-ray image $\hat{x}_p$. We apply $m_p$ for all $T$ timesteps, unlike Anatomy Controller, as our goal here is to infuse the detailed pathology, which requires iteration over entire $T$ steps. Algorithm \ref{alg:main_algo} outlines the inference process of LDM using anatomy and pathology controllers.
\section{Experiments} \label{sec:experiment}
\subsection{Dataset} 

\subsubsection{MaCheX}: The VAE models used in LDM, as well as the Anatomy Controller, were pre-trained on the Massive Chest X-ray (MaCheX) dataset \cite{weber2023cascaded}, which is a collection of chest X-ray images from different publicly available datasets. It contains 65,471 frontal AP/PA X-ray images collected from the designated train subsets of six large chest X-ray datasets, including ChestX-ray14 \cite{wang2017chestx}, CheXpert \cite{irvin2019chexpert}, MIMIC-CXR \cite{johnson2019mimic_web}, PadChest \cite{bustos2020padchest}, BRAX \cite{reis2022brax} and VinDR-CXR \cite{nguyen2022vindr} datasets. All the scans in MaCheX are rescaled so that the shortest edge meets a 1024-pixel resolution and are then center-cropped to $1024 \times 1024$ pixels. \\

\subsubsection{MIMIC-CXR}: Our text-to-image LDM backbone is trained on the X-ray image and text-label pairs of the MIMIC-CXR dataset \cite{johnson2019mimic}, which is a large collection of 377,110 chest X-ray images and corresponding free-text radiology reports and labels. During training, our model used over 120,000 Antero-Posterior (AP) view images from the training subset of the dataset, while we used the official test split to evaluate model performance. All the images were resized to $256 \times 256$ pixels before training and were randomly rotated $\pm 15$ degrees during the training. 

\subsection{Implementation Details} 
The VAE models within the LDM and the Anatomy Controller are trained using a downsampling factor of 4 and have the same architecture. For LDM's VAE, we used the pre-trained weights from \cite{weber2023cascaded}, further fine-tuned with 50 epochs for the Anatomy Controller. The text-to-image LDM model is trained on 256$\times$256 images for 100 epochs with a constant learning rate of $10^{-5}$ and batch size of 8. Furthermore, we used a linear $\beta$ noise scheduler with range $(0.0015, 0.0295)$ and set $s=50$ and $s=T$ for the Anatomy Controller and Pathology Control with $T=100$, respectively. Our experiments are conducted on two Nvidia RTX A6000 GPUs and implemented using the PyTorch \cite{paszke2019pytorch} framework.

\begin{figure}[t]
    \centering
    \includegraphics[width=0.5\textwidth]{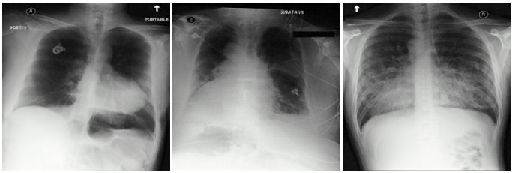}
    \caption{Images with unrealistic anatomical structures (i.e., heart at the wrong location) generated during one of the experiments can achieve a low FID score of $\sim$30. This supports our claim that the FID score does not provide any information about image realism.}
    \label{fig:bad_images}
\end{figure}

\subsection{Evaluation framework}

We used a combination of quantitative and qualitative metrics to evaluate the medical realism in the generated X-ray images. We aimed to quantify clinical realism by evaluating the models' ability to infuse correct pathology and generate images with realistic anatomy. \\

\subsubsection{Quantitative Evaluation} 

The quantitative performance evaluation of generative models typically includes assessing the fidelity of the generated data. For this, Fréchet Inception Distance (FID) \cite{heusel2017gans} is the most commonly used metric, which measures the performance of the generative model by comparing the distributions of generated and real datasets using an ImageNet-trained Inception model \cite{szegedy2015going}. However, while effective for natural images, this approach falls short in the medical domain, where clinical realism is highly important. Simply comparing distributions between datasets does not capture the features critical for medical images, rendering FID ineffective for medical applications. This limitation is clearly illustrated in Fig. \ref{fig:bad_images}, where images with significant artifacts achieve low FID, underscoring its shortcomings in assessing the aspects that are necessary to make medical data useful. Nonetheless, we report FID scores for all methods to show how FID changes with other metrics. However, relying on FID alone can be highly misleading, especially when the goal is to capture clinical realism. Addressing this, we suggest a more comprehensive evaluation framework, combining task-specific metrics that reflect the important aspects of medical imaging. This approach is necessary to ensure that generative models meet the specific requirements of medical applications. Conventional metrics like FID, while useful in broader contexts, can often miss important aspects when applied to medical imaging.

{\renewcommand{\arraystretch}{1.2}
\begin{table*}[t]
\centering
\caption{
Performance comparison of XReal with SOTA X-ray image generation methods using our quantitative and qualitative evaluation framework. What we refer to as the real images is the MIMIC-CXR test set to establish the upper bound of performance.
$^{\ddagger}$ControlNet \cite{zhang2023adding} was implemented using our LDM backbone.  $^{\dagger}$Reproduced results using the same data split trained on the MIMIC-CXR train set and tested on the test set.}

\resizebox{\textwidth}{!}{
\begin{tabular}{clccccccc}
\hline
\multirow{2}{*}{\textbf{Model type}} &  & \multicolumn{5}{c}{\textbf{Quantitative Results}} & \multicolumn{2}{c}{\textbf{Avg. Radiologist Scores}} \\ 
& \multirow{-2}{*}{\textbf{Model}} & MS-SSIM$\uparrow$ & FID$\downarrow$ & Dice$\uparrow$ & F1$\uparrow$ & AUC$\uparrow$ & Anatomy$\uparrow$ & Pathology$\uparrow$ \\

\hline

\multirow{2}{*}{\textbf{Text-to-Image}} & Cheff$^{\dagger}$ \cite{weber2023cascaded} & 0.415 & \textbf{24.640} & 0.500 & 0.510 & 0.640 & 2.927 & 3.180 \\ 

& RoentGen \cite{chambon2022roentgen} & 0.386 & 82.140 & 0.631 & 0.550 & \textbf{0.800} &  \underline{3.761} & 3.130 \\ 
\hline

\multirow{2}{*}{\textbf{Text + Spatial Control}} & ControlNet$^{\ddagger}$ \cite{zhang2023adding} & \underline{0.630} & \underline{29.480} &  \underline{0.835} & \underline{0.560} & 0.740 & 3.421 & \underline{3.372} \\ 

& \cellcolor[gray]{.9}XReal (Ours) & \cellcolor[gray]{.9}\textbf{0.701} & \cellcolor[gray]{.9}55.120 &  \cellcolor[gray]{.9}\textbf{0.838} & \cellcolor[gray]{.9}\textbf{0.570} & \cellcolor[gray]{.9}\underline{0.743} & \cellcolor[gray]{.9}\textbf{4.167} & \cellcolor[gray]{.9}\textbf{4.130} \\ 
\hline
\textbf{ } & \textcolor{gray}{Real Images} & \textcolor{gray}{---} & \textcolor{gray}{---} & \textcolor{gray}{---} & \textcolor{gray}{0.610} & \textcolor{gray}{0.800} & \textcolor{gray}{3.631} & \textcolor{gray}{3.561} \\  
\hline
\end{tabular}}
\label{tab:results}
\end{table*}}

We also used the Multi-Scale Structural Similarity Index (MS-SSIM) \cite{wang2003multiscale} to assess the realism of the generated data by comparing the real images with the generated ones. MS-SSIM evaluates the luminance, contrast, and structure of two images at multiple scales, providing a comprehensive assessment of variations at different levels of detail. While MS-SSIM is traditionally used to quantify diversity in generated data—where lower values indicate higher diversity—our goal was to measure image realism. To achieve this, we calculated the MS-SSIM between a real image and an image generated using the corresponding anatomy mask and pathology label from the same image. In this context, a higher MS-SSIM value indicates greater realism as it compares any structural inconsistencies (or artifacts) that are not considered by other metrics. Furthermore, to show the diversity offered by our method, we generated images using a variety of anatomy masks and applied different image transformations (e.g., rotation), as demonstrated in Fig. \ref{fig:result}. 

In addition to FID and MS-SSIM, we evaluated the models using a multi-label pathology classification task. For this, classification performance measures ($F_1$ and AUC) are calculated by passing the generated image through DenseNet-121 \cite{huang2017densely} model. This classification model is trained on the MIMIC-CXR dataset and performs comparably to the benchmark \cite{seyyed2020chexclusion} on the MIMIC-CXR test set. The classification metrics, particularly the $F_1$ score, are useful in imbalanced dataset settings and evaluate the presence and absence of the desired pathology in the generated image. In the absence of a dedicated pathology detection model, classification can offer indirect insights into spatial control over pathology localization. This is because a pathology is only considered correct if it appears in the appropriate region. As such, when a pathology is classified correctly in a specific area of the image, it can suggest that the disease has not only been detected but also manifested in the intended location. 

We used the Dice metric to evaluate the anatomical realism and spatial alignment between real and generated X-ray images. We focused on segmenting the lungs, heart, and aorta in real and generated images, as these organs are relevant to the thoracic pathologies of interest. Since MIMIC-CXR does not include segmentation masks, we obtain these masks using a pre-trained X-ray segmentation model available in TorchXRayVision library \cite{cohen2022torchxrayvision}. These masks were used as pseudo-labels to train the Anatomy Controller and ControlNet \cite{zhang2023adding} and to evaluate the performance of both models on the MIMIC-CXR test set. The Dice score for text-to-image models indicates the average overlap by chance between an original X-ray image organs and the image generated using the corresponding text report. \\

\subsubsection{Qualitative Evaluation}

Another important aspect of assessing the performance of a generative model is to do a visual or qualitative evaluation. This is particularly relevant to the medical domain, where it is very difficult to quantify the realism of the generated data via other metrics. In this work, two experienced radiologists conducted the qualitative evaluation in a blind review setting. We generated the images using all four methods (including XReal) by providing the corresponding reports, pathology labels, and anatomy mask of the MIMIC-CXR test set to the models. Both radiologists were asked to rank the generated and real images independently, from 1 (lowest) to 5 (highest), based on anatomy, pathology realism, and image quality. In conjunction with the quantitative metrics, this clinically driven evaluation allowed us to compare medical realism in the generated X-ray images and draw reliable results and meaningful conclusions.
\section{Results} \label{sec:results}

\subsection{Quantitative Results}

Table \ref{tab:results} summarizes XReal's quantitative results and compares them with state-of-the-art (SOTA) image generation methods. XReal achieves the highest MS-SSIM score by a significant margin, demonstrating its ability to generate realistic and cleaner peripheries (the region outside the lungs, heart, and aorta). In comparison, the MS-SSIM scores for text-to-image models serve as a baseline, offering insights into the structural similarity between real and generated images in the absence of spatial control. We also report the FID score calculated for each method. Cheff \cite{weber2023cascaded} and ControlNet \cite{zhang2023adding} achieve a lower FID score, followed by our method. The FID score itself does not provide any information about medical realism, as it only compares the distribution of real and generated data. Our results also show that there is no correlation between the FID score and any other quantitative or qualitative metrics. Moreover, some of the generated images with easily visible artifacts and noise achieve significantly lower FID, as shown in Fig. \ref{fig:bad_images}, highlighting the shortcoming of FID as a metric for medical imaging. This strengthens our initial conjecture that relying on the FID scores alone can be highly misleading, particularly in the medical domain, where image realism significantly impacts the usefulness of the generated data. 

Classification performance measures are also included in Table \ref{tab:results}, where $F_1$ score and AUC are calculated using the DenseNet-121 model \cite{huang2017densely} trained on the MIMIC-CXR dataset. We compare our model with SOTA text-to-image diffusion models and existing controllable diffusion model, ControlNet \cite{zhang2023adding}. The reported macro-$F_1$ score is calculated by aggregating the $F_1$ scores for each class, making it particularly suitable for imbalanced datasets. The $F_1$ score achieved by XReal outperforms other methods while also achieving the second-highest AUC. This demonstrates that XReal effectively introduces the specified pathology in the generated image while allowing precise control over its location. Other methods not only achieve lower classification scores but also lack the spatial control offered by XReal. Despite the increased complexity and precision required, XReal outperforms these methods, making it the only approach that offers spatial control over pathology manifestation.

We evaluate the spatial control offered by each model using a pre-trained chest X-ray segmentation model \cite{cohen2022torchxrayvision}. The goal is to check the overlap between the organs in the generated image and the real image associated with the anatomy mask. To this end, we compare our model with ControlNet \cite{zhang2023adding}, trained on identical data splits. We calculate the Dice score between segmented lungs, aorta, and heart from the original and generated images. Table \ref{tab:results} shows that XReal outperforms ControlNet by offering better anatomical control using only 55M parameters compared to ControlNet's 217M parameters (excluding LDM's parameters in both models). Furthermore, our model requires only a single pass through the Anatomy Controller, compared to $T$ ($\approx100$) iterations for ControlNet's encoder. The dice score for text-to-image models shows the average overlap between the image associated with the input report and the output images using solely textual input.

\subsection{Qualitative Results}

To solidify our assessment of the generated X-rays' clinical realism, two expert radiologists reviewed the images in a blind review setting. As shown in Table \ref{tab:results}, XReal outperformed other methods by a large margin with scores of 4.167 and 4.130 for anatomy and pathology, respectively, while surpassing real images in both anatomy (+0.536) and pathology realism (+0.569) evaluation. This improvement, particularly over real images, can be attributed to a number of factors, including XReal's ability to accurately generate all anatomical structures based on the provided anatomical mask ($m_a$), making both the anatomy and pathology clearer. Additionally, since the anatomical mask was obtained using a segmentation model, XReal can potentially avoid including artifacts that might be present in the real images but are not captured in the mask. This can lead to clearer anatomy and more evident pathology manifestations. However, it would be important to investigate incorporating X-ray artifacts in the future. 

Our results also suggest that quantifying medical realism is a highly challenging task. Therefore, it is very important to use a combination of metrics that target different aspects of realism. Furthermore, augmenting quantitative results with human expert evaluation is crucial for comparing the methods. 
\begin{figure}[t]
    \centering
    \includegraphics[width=\linewidth]{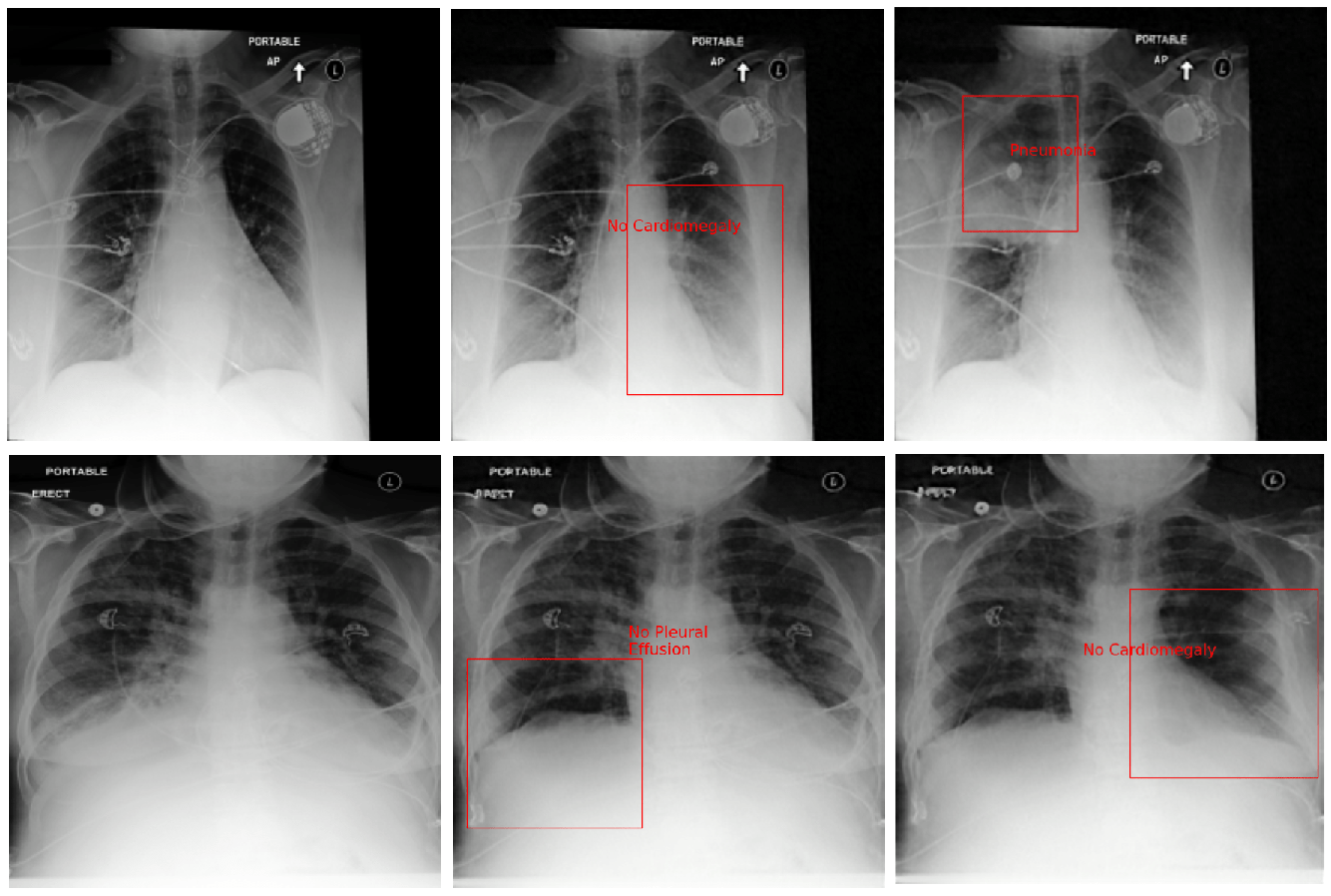}
    \caption{
    In each row, we show a sample X-ray image with an existing pathology (Left), where we use XReal to remove the pathology (Center) and then add a different pathology (Right). 
    }
    \label{fig:path_add_remove}
\end{figure}

\begin{figure}[t]
    \centering
    \includegraphics[width=0.5\textwidth]{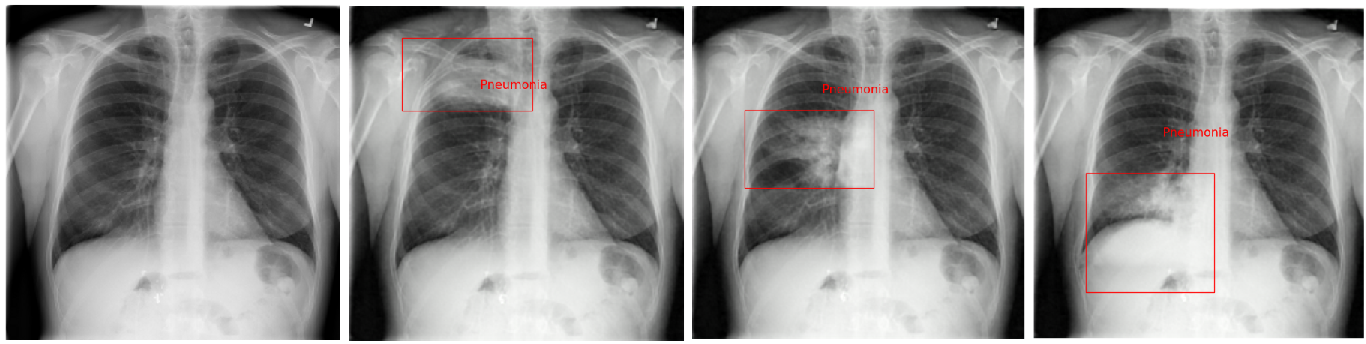}
    \caption{A sample X-ray image along with three example generated images using XReal, where the pathology location is moved vertically along the right lung.}
    \label{fig:discussion_fig}
\end{figure}

\begin{figure}[t]
    \centering
    \includegraphics[width=0.5\textwidth]{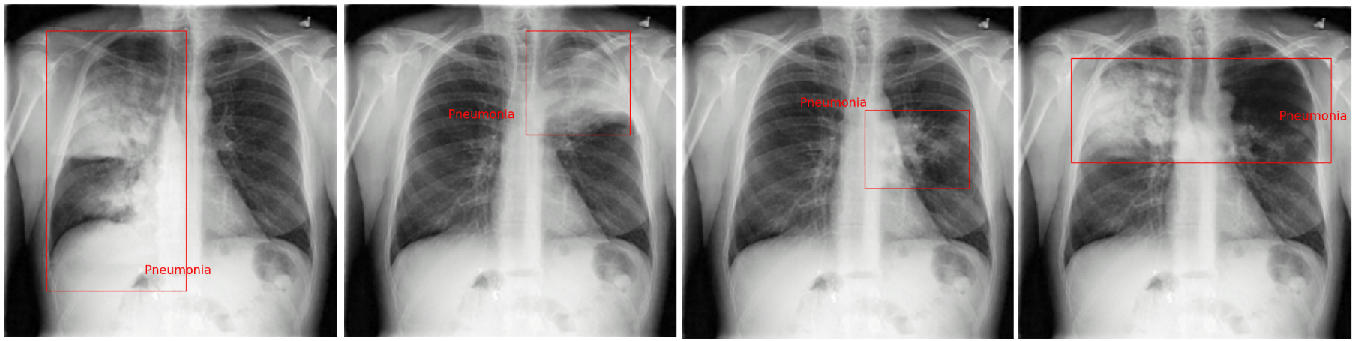}
    \caption{
    The same anatomical structure was generated on different X-ray images with different pneumonia severities using XReal.
    }
    \label{fig:path_location_control}
\end{figure}

\begin{figure}[t]
    \centering
    \includegraphics[width=0.5\textwidth]{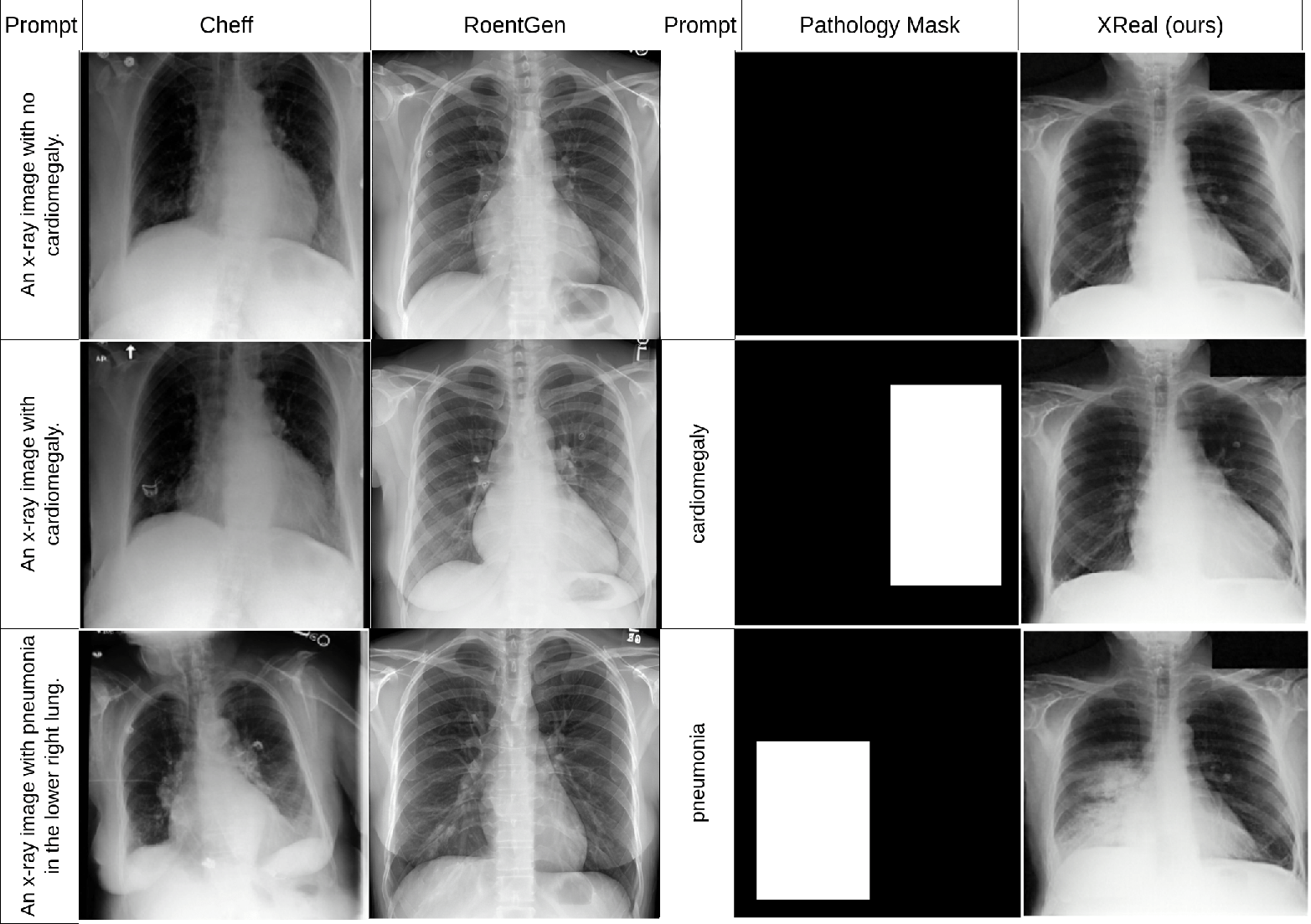}
    \caption{Pathology infusion in a specific location using different diffusion models. Text-to-image methods (Cheff and RoentGen) fail to localize the specified pathology in the right location and struggle to incorporate the given prompt precisely. On the other hand, XReal can generate an X-ray image with a given anatomy \textit{and} seamlessly infuse the specified lesion at the desired location.}
    \label{fig:pathology_infusion}
\end{figure}

\begin{figure}[t]
    \centering
    \includegraphics[width=0.5\textwidth]{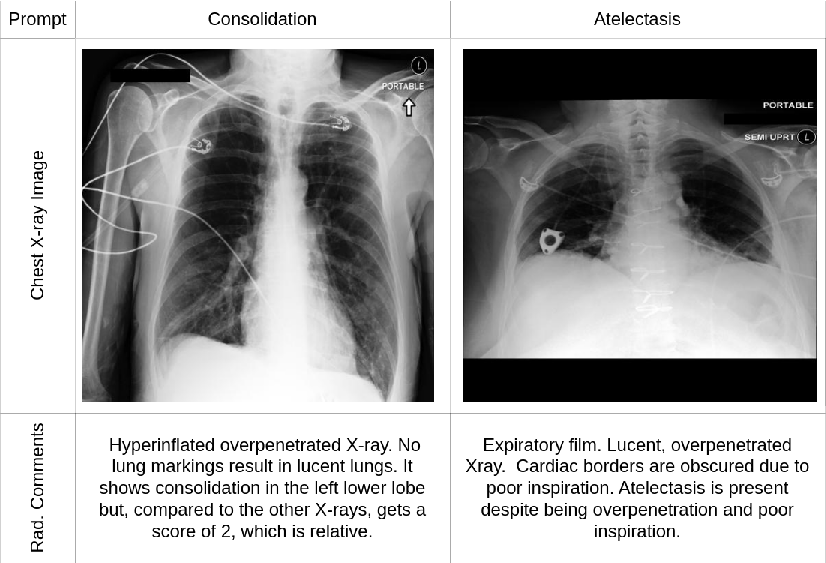}
    \caption{Sample explanation by a radiologist for the lower ranking of real images. These real images show the mentioned pathology; however, they may contain some anatomical artifacts (e.g., obscured cardiac borders), and the pathology is not clearly manifested compared to the other images (not shown in this figure).}
    \label{fig:rad_explainer}
\end{figure}

\section{Discussion} \label{sec:discussion}

XReal addresses the issue of unrealistic medical image generation by introducing control over the anatomy or pathology in diffusion models. We compared our model with text-to-image and controllable diffusion models. Previously, ControlNet \cite{zhang2023adding} has been used in natural images, but no attempts were made to train it for X-ray generation. Although ControlNet \cite{zhang2023adding} offers comparable anatomical control, its diffusion-based mask encoder uses approximately four times more parameters than XReal's VAE-based Anatomy Controller while requiring $\sim 100 \times$ more iterations. Furthermore, ControlNet does not provide spatial control over the pathology, making it susceptible to the issue of textual ambiguities faced by the text-to-image generative models. 


XReal's ability to control the location of pathology and anatomy makes it particularly useful for various clinical applications. The Pathology Controller in XReal can be used for image editing by adding or removing a particular disease from a given X-ray image. Fig. \ref{fig:path_add_remove} shows the removal and addition of different pathologies from a sample X-ray image while keeping the original anatomy intact and without introducing any artifacts. Additionally, Fig. \ref{fig:discussion_fig} shows the introduction of pathology to different locations in a sample X-ray image, demonstrating the translation of a single pathology to multiple positions in the lungs. Furthermore, the intensity or size of the disease manifestation can also be controlled by changing the size of the bounding box in $m_p$ (Fig. \ref{fig:path_location_control}). This level of control is clearly lacking in text-to-image models, as shown in Fig. \ref{fig:pathology_infusion}. 

XReal's image editing abilities have clinical significance, as it can generate different cases from a given X-ray image (or patient). This can be particularly useful for counterfactual image generation \cite{cohen2021gifsplanation,gu2023biomedjourney} and disease prognosis studies, where such images can be used to answer \textit{what-if} questions related to pulmonary diseases. Other applications of XReal could be in training radiologists, patient education, and simulation software for pulmonary diseases, where XReal can be used to generate different cases for a given anatomy.

We also introduce a comprehensive evaluation framework to assess clinical realism, which includes classification, segmentation metrics, and expert human evaluation. Traditional metrics like FID can be misleading in the medical domain, as they focus solely on data distribution. Medical images are far more complex than natural images, containing intricate details that require more nuanced evaluation. Images with similar distributions can have significant differences in terms of disease impressions and diagnosis. This issue is evident in Fig. \ref{fig:bad_images} as well and highlights a shortcoming of natural imaging domain metrics in the medical domain. Furthermore, we used the classification metrics to quantify the manifestation of pathology. The F1 score takes into account both false positive and false negative cases and, therefore, provides a better performance measure for imbalance data. The classification score does not account for the pathology's location. However, a pathology is only considered correct when it appears in the appropriate region (e.g., cardiomegaly in the heart or pneumonia in the lungs). Therefore, a correct pathology classification, when it is introduced in a specific location within the image, suggests that the disease has been accurately manifested in the intended location. A better way could have been to use a pathology detection model; however, there is no publicly available pre-trained detection model or dataset with bounding box annotations that covers the disease labels in the used dataset. Besides classification metrics, the dice score shows the anatomical alignment between the input mask and the generated image. Hence, the combined classification and segmentation metrics provide a multifaceted assessment of clinical realism.

The radiologists' evaluation shows that XReal outperforms all other methods by a significant margin and also achieves a better image realism score than the real images. The higher scores than the real images can be due to a number of factors: (1) XReal does not introduce the X-ray artifacts that are often visible in real X-ray scans but are not captured in the input mask $m_a$. This can lead to the generation of cleaner X-ray images that seem more realistic for radiologists. (2) The MIMIC dataset labels are not guaranteed to be accurate and may contain false positives/negatives. These labels are extracted from the clinical reports using the CheXpert-labeler \cite{irvin2019chexpert} and contain 1, 0, and -1 labels, where -1 indicates an uncertain presence or absence of a pathology. The automated labelers can possibly introduce discrepancies in labels and clinical reports, as discussed in \cite{olatunji2019caveats}. Moreover, we changed the label -1 to 0, which can further increase the mismatch with the real image. (3) The human expert rankings are not absolute, and a relative ranking of generated and real images means that real images can get relatively lower scores despite containing the correct pathology. In this case, XReal generates an image with relatively clearer features that are more realistic and better spotted by the experts, leading to better ranking. (4) The image quality can also affect anatomical realism (e.g., it can obscure anatomical structures or blackout vascular markings) and disease manifestation (resulting in false positives or negatives); hence, any artifacts or markings that affect the image quality can lead to a lower score for pathology realism and vice-versa. 

To understand this effect further, we asked the radiologists to re-evaluate a subset of the images and share their comments on why the real images are ranked lower than the generated images. Fig. \ref{fig:rad_explainer} shows a sample explanation for the ranking by one of the radiologists. Their explanation suggests that although the real images have the mentioned pathology ($p$), they may contain anatomical artifacts such as obscured cardiac borders or relatively unclear pathology manifestation compared to other images, leading to a lower ranking.

Another experiment was conducted to evaluate the choice of using a label-to-image LDM in XReal instead of a report-to-image diffusion model. For this, we compared the label-to-image LDM in XReal with the Cheff \cite{weber2023cascaded} and Roentgen \cite{chambon2022roentgen} report-to-image diffusion models. Cheff and Roentgen were trained on paired radiology reports and images from the MIMIC dataset, while our LDM was trained on paired pathology labels and X-ray images. The results show that the label-to-image LDM outperformed the report-to-image models, achieving the highest $F_1$ score of 0.59 and an AUC score of 0.78 (second only to Roentgen \cite{chambon2022roentgen}). This suggests that longer prompts or reports do not necessarily improve pathology manifestation. While radiology reports provide more detailed prompts, they may not enhance accuracy in these models. These findings support our decision to use a label-to-image LDM in XReal, where text provides only pathology label $p$, and all spatial information is conveyed through segmentation masks $m_a$ and $m_p$.

\section{Limitations and Future Work}
Our method has a number of limitations that can be addressed in future works. XReal uses two stages for anatomy control and pathology infusion, which can be improved by combining the Anatomy and Pathology Controllers in XReal. Another limitation could be the possible removal of important artifacts from the generated X-ray image. These artifacts could be useful in generating images to mimic real-life scenarios with more details and to study their effect on the radiologists' evaluation. In the future, it will be interesting to have a method to control the different types of artifacts and study their effect on clinical realism. Moreover, our proposed method could be used for other modalities where having control over anatomy and pathology has clinical significance.

\section{Conclusion} \label{sec:conclusion}
We introduce XReal, a controllable diffusion model for realistic chest X-ray image generation through precise control over anatomy and pathology. We compare the medical image realism in the generated X-ray images via a combination of different metrics. XReal provides control over the anatomy of a generated X-ray image using a free-form anatomy mask. It can also be used to add or remove different pathologies from a given X-ray image, which can have various clinical applications. In the future, realistic medical image generation can be further explored to develop methods that generate useful medical data. It is also important to understand more aspects of clinical realism and to develop better metrics tailored for medical applications.

\bibliographystyle{IEEEtran} 
\bibliography{references}

\begin{thebibliography}{10}
\providecommand{\url}[1]{#1}
\csname url@samestyle\endcsname
\providecommand{\newblock}{\relax}
\providecommand{\bibinfo}[2]{#2}
\providecommand{\BIBentrySTDinterwordspacing}{\spaceskip=0pt\relax}
\providecommand{\BIBentryALTinterwordstretchfactor}{4}
\providecommand{\BIBentryALTinterwordspacing}{\spaceskip=\fontdimen2\font plus
\BIBentryALTinterwordstretchfactor\fontdimen3\font minus
  \fontdimen4\font\relax}
\providecommand{\BIBforeignlanguage}[2]{{%
\expandafter\ifx\csname l@#1\endcsname\relax
\typeout{** WARNING: IEEEtran.bst: No hyphenation pattern has been}%
\typeout{** loaded for the language `#1'. Using the pattern for}%
\typeout{** the default language instead.}%
\else
\language=\csname l@#1\endcsname
\fi
#2}}
\providecommand{\BIBdecl}{\relax}
\BIBdecl

\bibitem{chambon2022roentgen}
P.~Chambon, C.~Bluethgen, J.-B. Delbrouck \emph{et~al.}, ``Roentgen:
  vision-language foundation model for chest x-ray generation,'' \emph{arXiv
  preprint arXiv:2211.12737}, 2022.

\bibitem{weber2023cascaded}
T.~Weber, M.~Ingrisch, B.~Bischl, and D.~R{\"u}gamer, ``Cascaded latent
  diffusion models for high-resolution chest x-ray synthesis,'' in
  \emph{Pacific-Asia Conference on Knowledge Discovery and Data Mining}.\hskip
  1em plus 0.5em minus 0.4em\relax Springer, 2023, pp. 180--191.

\bibitem{ramesh2022hierarchical}
A.~Ramesh, P.~Dhariwal, A.~Nichol \emph{et~al.}, ``Hierarchical
  text-conditional image generation with clip latents,'' \emph{arXiv preprint
  arXiv:2204.06125}, 2022.

\bibitem{brown2020language}
T.~Brown, B.~Mann, N.~Ryder \emph{et~al.}, ``Language models are few-shot
  learners,'' \emph{Advances in neural information processing systems},
  vol.~33, 2020.

\bibitem{rombach2022high}
R.~Rombach, A.~Blattmann, D.~Lorenz \emph{et~al.}, ``High-resolution image
  synthesis with latent diffusion models,'' in \emph{Proceedings of the
  IEEE/CVF CVPR}, 2022.

\bibitem{rawte2023survey}
V.~Rawte, A.~Sheth, and A.~Das, ``A survey of hallucination in large foundation
  models,'' \emph{arXiv preprint arXiv:2309.05922}, 2023.

\bibitem{avrahami2023spatext}
O.~Avrahami, T.~Hayes, O.~Gafni \emph{et~al.}, ``Spatext: Spatio-textual
  representation for controllable image generation,'' in \emph{Proceedings of
  the IEEE/CVF Conference on Computer Vision and Pattern Recognition}, 2023,
  pp. 18\,370--18\,380.

\bibitem{casanova2023controllable}
A.~Casanova, M.~Careil, A.~Romero-Soriano \emph{et~al.}, ``Controllable image
  generation via collage representations,'' \emph{arXiv preprint
  arXiv:2304.13722}, 2023.

\bibitem{kingma2013auto}
D.~P. Kingma and M.~Welling, ``Auto-encoding variational bayes,'' \emph{arXiv
  preprint arXiv:1312.6114}, 2013.

\bibitem{goodfellow2014generative}
I.~Goodfellow, J.~Pouget-Abadie, M.~Mirza \emph{et~al.}, ``Generative
  adversarial nets,'' \emph{Advances in neural information processing systems},
  vol.~27, 2014.

\bibitem{dhariwal2021diffusion}
P.~Dhariwal and A.~Nichol, ``Diffusion models beat gans on image synthesis,''
  \emph{Advances in neural information processing systems}, vol.~34, 2021.

\bibitem{gu2023biomedjourney}
Y.~Gu, J.~Yang, N.~Usuyama \emph{et~al.}, ``Biomedjourney: Counterfactual
  biomedical image generation by instruction-learning from multimodal patient
  journeys,'' \emph{arXiv preprint arXiv:2310.10765}, 2023.

\bibitem{segal2021evaluating}
B.~Segal, D.~M. Rubin, G.~Rubin, and A.~Pantanowitz, ``Evaluating the clinical
  realism of synthetic chest x-rays generated using progressively growing
  gans,'' \emph{SN Computer Science}, vol.~2, no.~4, p. 321, 2021.

\bibitem{karras2017progressive}
T.~Karras, T.~Aila, S.~Laine, and J.~Lehtinen, ``Progressive growing of gans
  for improved quality, stability, and variation,'' \emph{arXiv preprint
  arXiv:1710.10196}, 2017.

\bibitem{ng2023generative}
M.~F. Ng and C.~A. Hargreaves, ``Generative adversarial networks for the
  synthesis of chest x-ray images,'' \emph{Engineering Proceedings}, vol.~31,
  no.~1, p.~84, 2023.

\bibitem{radford2015unsupervised}
A.~Radford, L.~Metz, and S.~Chintala, ``Unsupervised representation learning
  with deep convolutional generative adversarial networks,'' \emph{arXiv
  preprint arXiv:1511.06434}, 2015.

\bibitem{gulrajani2017improved}
I.~Gulrajani, F.~Ahmed, M.~Arjovsky \emph{et~al.}, ``Improved training of
  wasserstein gans,'' \emph{Advances in neural information processing systems},
  2017.

\bibitem{ciano2021multi}
G.~Ciano, P.~Andreini, T.~Mazzierli \emph{et~al.}, ``A multi-stage gan for
  multi-organ chest x-ray image generation and segmentation,''
  \emph{Mathematics}, 2021.

\bibitem{yang2020xraygan}
X.~Yang, N.~Gireesh, E.~Xing \emph{et~al.}, ``Xraygan: Consistency-preserving
  generation of x-ray images from radiology reports,'' \emph{arXiv preprint
  arXiv:2006.10552}, 2020.

\bibitem{chambon2022adapting}
P.~Chambon, Bluethgen \emph{et~al.}, ``Adapting pretrained vision-language
  foundational models to medical imaging domains,'' \emph{arXiv preprint
  arXiv:2210.04133}, 2022.

\bibitem{radford2021learning}
A.~Radford, J.~W. Kim, C.~Hallacy \emph{et~al.}, ``Learning transferable visual
  models from natural language supervision,'' in \emph{International conference
  on machine learning}.\hskip 1em plus 0.5em minus 0.4em\relax PMLR, 2021.

\bibitem{packhauser2023generation}
K.~Packh{\"a}user and o.~Folle, ``Generation of anonymous chest radiographs
  using latent diffusion models for training thoracic abnormality
  classification systems,'' in \emph{2023 IEEE 20th International Symposium on
  Biomedical Imaging (ISBI)}.\hskip 1em plus 0.5em minus 0.4em\relax IEEE,
  2023.

\bibitem{liu2023textdiff}
B.~Liu, Z.~Yang, P.~Wang \emph{et~al.}, ``Textdiff: Mask-guided residual
  diffusion models for scene text image super-resolution,'' \emph{arXiv
  preprint arXiv:2308.06743}, 2023.

\bibitem{meng2022sdedit}
C.~Meng \emph{et~al.}, ``{SDE}dit: Guided image synthesis and editing with
  stochastic differential equations,'' in \emph{International Conference on
  Learning Representations}, 2022.

\bibitem{xie2023smartbrush}
S.~Xie, Z.~Zhang, Z.~Lin \emph{et~al.}, ``Smartbrush: Text and shape guided
  object inpainting with diffusion model,'' in \emph{Proceedings of the
  IEEE/CVF Conference on Computer Vision and Pattern Recognition}, 2023.

\bibitem{van2023echocardiography}
P.~N. Van, D.~T. Minh, H.~P. Huy \emph{et~al.}, ``Echocardiography video
  synthesis from end diastolic semantic map via diffusion model,'' \emph{arXiv
  preprint arXiv:2310.07131}, 2023.

\bibitem{dorjsembe2023conditional}
Z.~Dorjsembe, H.-K. Pao, S.~Odonchimed, and F.~Xiao, ``Conditional diffusion
  models for semantic 3d medical image synthesis,'' \emph{arXiv preprint
  arXiv:2305.18453}, 2023.

\bibitem{wu2023harnessing}
Q.~Wu, Y.~Liu, H.~Zhao, T.~Bui, Z.~Lin, Y.~Zhang, and S.~Chang, ``Harnessing
  the spatial-temporal attention of diffusion models for high-fidelity
  text-to-image synthesis,'' in \emph{Proceedings of the IEEE/CVF International
  Conference on Computer Vision}, 2023, pp. 7766--7776.

\bibitem{couairon2023zero}
G.~Couairon, M.~Careil, M.~Cord \emph{et~al.}, ``Zero-shot spatial layout
  conditioning for text-to-image diffusion models,'' in \emph{Proceedings of
  the IEEE/CVF International Conference on Computer Vision}, 2023.

\bibitem{park2022shape}
D.~H. Park, G.~Luo, C.~Toste, S.~Azadi, X.~Liu, M.~Karalashvili, A.~Rohrbach,
  and T.~Darrell, ``Shape-guided diffusion with inside-outside attention,''
  \emph{arXiv preprint arXiv:2212.00210}, 2022.

\bibitem{xie2023boxdiff}
J.~Xie, Y.~Li, Y.~Huang \emph{et~al.}, ``Boxdiff: Text-to-image synthesis with
  training-free box-constrained diffusion,'' in \emph{Proceedings of the
  IEEE/CVF International Conference on Computer Vision}, 2023.

\bibitem{zhang2023adding}
o.~Zhang, Lvmin, ``Adding conditional control to text-to-image diffusion
  models,'' in \emph{Proceedings of the IEEE/CVF International Conference on
  Computer Vision}, 2023.

\bibitem{ratzlaff2019hypergan}
N.~Ratzlaff and L.~Fuxin, ``Hypergan: A generative model for diverse,
  performant neural networks,'' in \emph{International Conference on Machine
  Learning}.\hskip 1em plus 0.5em minus 0.4em\relax PMLR, 2019, pp. 5361--5369.

\bibitem{kumar2023coronetgan}
A.~Kumar, K.~Anand, S.~Mandloi \emph{et~al.}, ``Coronetgan: Controlled pruning
  of gans via hypernetworks,'' in \emph{Proceedings of the IEEE/CVF ICCV},
  2023.

\bibitem{ho2020denoising}
J.~Ho, A.~Jain, and P.~Abbeel, ``Denoising diffusion probabilistic models,''
  \emph{Advances in neural information processing systems}, vol.~33, 2020.

\bibitem{wang2017chestx}
X.~Wang, Y.~Peng, L.~Lu, Z.~Lu, M.~Bagheri, and R.~M. Summers, ``Chestx-ray8:
  Hospital-scale chest x-ray database and benchmarks on weakly-supervised
  classification and localization of common thorax diseases,'' in
  \emph{Proceedings of the IEEE conference on computer vision and pattern
  recognition}, 2017, pp. 2097--2106.

\bibitem{irvin2019chexpert}
J.~Irvin, P.~Rajpurkar, M.~Ko, Y.~Yu, S.~Ciurea-Ilcus, C.~Chute, H.~Marklund,
  B.~Haghgoo, R.~Ball, K.~Shpanskaya \emph{et~al.}, ``Chexpert: A large chest
  radiograph dataset with uncertainty labels and expert comparison,'' in
  \emph{Proceedings of the AAAI conference on artificial intelligence},
  vol.~33, no.~01, 2019, pp. 590--597.

\bibitem{johnson2019mimic_web}
A.~E. Johnson, T.~J. Pollard, S.~J. Berkowitz \emph{et~al.}, ``Mimic-cxr, a
  de-identified publicly available database of chest radiographs with free-text
  reports,'' \emph{Scientific Data}, vol.~6, no.~1, p. 317, 2019.

\bibitem{bustos2020padchest}
A.~Bustos, A.~Pertusa, J.-M. Salinas, and M.~De~La Iglesia-Vaya, ``Padchest: A
  large chest x-ray image dataset with multi-label annotated reports,''
  \emph{Medical image analysis}, vol.~66, p. 101797, 2020.

\bibitem{reis2022brax}
E.~P. Reis, J.~P. De~Paiva, M.~C. Da~Silva, G.~A. Ribeiro, V.~F. Paiva,
  L.~Bulgarelli, H.~M. Lee, P.~V. Santos, V.~M. Brito, L.~T. Amaral
  \emph{et~al.}, ``Brax, brazilian labeled chest x-ray dataset,''
  \emph{Scientific Data}, vol.~9, no.~1, p. 487, 2022.

\bibitem{nguyen2022vindr}
H.~Q. Nguyen, K.~Lam, L.~T. Le \emph{et~al.}, ``Vindr-cxr: An open dataset of
  chest x-rays with radiologist’s annotations,'' \emph{Scientific Data},
  vol.~9, no.~1, 2022.

\bibitem{johnson2019mimic}
A.~Johnson, T.~Pollard, R.~Mark \emph{et~al.}, ``Mimic-cxr database (version
  2.0. 0). physionet,'' 2019.

\bibitem{paszke2019pytorch}
A.~Paszke, S.~Gross, F.~Massa \emph{et~al.}, ``Pytorch: An imperative style,
  high-performance deep learning library,'' \emph{Advances in neural
  information processing systems}, vol.~32, 2019.

\bibitem{heusel2017gans}
M.~Heusel, H.~Ramsauer, T.~Unterthiner \emph{et~al.}, ``Gans trained by a two
  time-scale update rule converge to a local nash equilibrium,'' \emph{Advances
  in neural information processing systems}, vol.~30, 2017.

\bibitem{szegedy2015going}
C.~Szegedy, W.~Liu, Y.~Jia, P.~Sermanet, S.~Reed, D.~Anguelov, D.~Erhan,
  V.~Vanhoucke, and A.~Rabinovich, ``Going deeper with convolutions,'' in
  \emph{Proceedings of the IEEE conference on computer vision and pattern
  recognition}, 2015, pp. 1--9.

\bibitem{wang2003multiscale}
Z.~Wang, E.~P. Simoncelli, and A.~C. Bovik, ``Multiscale structural similarity
  for image quality assessment,'' in \emph{The Thrity-Seventh Asilomar
  Conference on Signals, Systems \& Computers, 2003}, vol.~2.\hskip 1em plus
  0.5em minus 0.4em\relax Ieee, 2003, pp. 1398--1402.

\bibitem{huang2017densely}
G.~Huang and o.~Liu, ``Densely connected convolutional networks,'' in
  \emph{Proceedings of the IEEE conference on computer vision and pattern
  recognition}, 2017.

\bibitem{seyyed2020chexclusion}
L.~Seyyed-Kalantari, G.~Liu, M.~McDermott, I.~Y. Chen, and M.~Ghassemi,
  ``Chexclusion: Fairness gaps in deep chest x-ray classifiers,'' in
  \emph{BIOCOMPUTING 2021: proceedings of the Pacific symposium}.\hskip 1em
  plus 0.5em minus 0.4em\relax World Scientific, 2020, pp. 232--243.

\bibitem{cohen2022torchxrayvision}
J.~P. Cohen \emph{et~al.}, ``Torchxrayvision: A library of chest x-ray datasets
  and models,'' in \emph{International Conference on Medical Imaging with Deep
  Learning}, 2022.

\bibitem{cohen2021gifsplanation}
J.~P. Cohen, R.~Brooks, S.~En \emph{et~al.}, ``Gifsplanation via latent shift:
  a simple autoencoder approach to counterfactual generation for chest
  x-rays,'' in \emph{Medical Imaging with Deep Learning}.\hskip 1em plus 0.5em
  minus 0.4em\relax PMLR, 2021.

\bibitem{olatunji2019caveats}
T.~Olatunji, L.~Yao, B.~Covington, A.~Rhodes, and A.~Upton, ``Caveats in
  generating medical imaging labels from radiology reports,'' \emph{arXiv
  preprint arXiv:1905.02283}, 2019.

\end{thebibliography}

\end{document}